\documentclass[twocolumn,floatfix,a4paper,showpacs,showkeys,nofootinbib,reprint,prl]{revtex4-1}
\usepackage{epsfig}
\usepackage{latexsym}
\usepackage{xspace}
\usepackage[colorlinks=true,linktocpage=true,linkcolor=blue,citecolor=blue,allcolors=blue]{hyperref}
\usepackage{url}
\usepackage[utf8]{inputenc}
\usepackage{indentfirst}
\usepackage{enumerate}
\usepackage{color}

\usepackage[caption=false,position=top]{subfig}

\usepackage{amsmath}
\usepackage{amssymb}

\usepackage[english]{babel}
\usepackage{url}

\AtBeginDocument{\renewcommand{\natexlab}[1]{#1}}

\newcommand{\mean}[1]{\langle #1 \rangle}

\newcommand{\npiNN}{\mean{n_\pi^{N\bar{N}}}}

\newcommand{\Nch}{d N_{\rm ch} / d \eta}
\newcommand{\sNN}{\sqrt{s_{\rm NN}}}

\newcommand{\eq}[1]{\begin{align} #1 \end{align}}

\begin{document}

\title{
Centrality dependence of proton and light nuclei yields as a consequence of baryon annihilation in the hadronic phase
}

\author{Volodymyr Vovchenko}
\affiliation{Institute for Nuclear Theory, University of Washington, Box 351550, Seattle, WA 98195, USA}
\affiliation{Nuclear Science Division, Lawrence Berkeley National Laboratory, 1 Cyclotron Road, Berkeley, CA 94720, USA}
\affiliation{Frankfurt Institute for Advanced Studies, Giersch Science Center,
D-60438 Frankfurt am Main, Germany}

\author{Volker Koch}
\affiliation{Nuclear Science Division, Lawrence Berkeley National Laboratory, 1 Cyclotron Road, Berkeley, CA 94720, USA}

\begin{abstract}
The centrality dependence of the $p/\pi$ ratio measured by the ALICE Collaboration in 5.02 TeV Pb-Pb collisions indicates a statistically significant suppression with the increase of the charged particle multiplicity once the centrality-correlated part of the systematic uncertainty is eliminated from the data.
We argue that this behavior can be attributed to baryon annihilation in the hadronic phase.
By implementing the $B\bar{B} \leftrightarrow 5\pi$ reaction within a generalized partial chemical equilibrium framework, we estimate the annihilation freeze-out temperature at different centralities, which decreases with increasing charged particle multiplicity and yields $T_{\rm ann} = 132 \pm 5$~MeV in 0-5\% most central collisions. This value is considerably below the hadronization temperature of $T_{\rm had} \sim 160$~MeV but above the thermal (kinetic) freeze-out temperature of $T_{\rm kin} \sim 100$~MeV.
Baryon annihilation reactions thus remain relevant in the initial stage of the hadronic phase but freeze out before (pseudo-)elastic hadronic scatterings. 
One experimentally testable consequence of this picture is a suppression of various light nuclei to proton ratios in central collisions of heavy ions.
\end{abstract}

\maketitle

\paragraph{\bf Introduction.}

Baryon-antibaryon annihilation is among the most important reactions in hadronic matter.
These reactions are responsible for the disappearance of antimatter during the expansion and cooling of the matter created in the Big Bang below the QCD transition temperature $T \lesssim 160$~MeV.
Conditions similar to the early Universe are recreated in little bangs -- relativistic heavy-ion collisions -- where baryon annihilation should play a significant role in the hadronic phase~\cite{Bass:2000ib}. 
Monte Carlo hadronic afterburners such as UrQMD~\cite{Bass:1998ca,Bleicher:1999xi} or SMASH~\cite{Weil:2016zrk} do predict sizable suppression of (anti)baryons yields due to the annihilations~\cite{Karpenko:2012yf,Becattini:2012sq,Steinheimer:2012rd,Becattini:2012xb,Steinheimer:2017vju,Oliinychenko:2018ugs}.

The suppression of the proton yield in central Pb-Pb collisions at the LHC relative to statistical hadronization model~(SHM) predictions~\cite{Abelev:2013vea} has been discussed as possible experimental evidence for baryon annihilation in the hadronic phase~\cite{Becattini:2012sq,Steinheimer:2012rd}.
However, it has also been pointed out that there are sizable theoretical uncertainties in SHM predictions of proton abundances due to modeling of meson-baryon interactions~\cite{Alba:2016hwx,Vovchenko:2018fmh,Andronic:2018qqt}, which could potentially explain the discrepancy, at least partially.
Furthermore, proton suppression due to baryon annihilation has been predicted based on transport model simulations of the hadronic phase that incorporate direct reactions such as $B+\bar{B} \to n \,\pi$~\cite{Cassing:1999es, Lin:2004en}, where typically $n \sim 5$~\cite{Dover:1992vj}, but not regeneration reactions, $n \pi \to B + \bar{B}$, thus violating detailed balance.
Implementation of multi-particle baryon regeneration reactions in transport codes is challenging~\cite{Garcia-Montero:2021haa} and if done properly can mitigate the effect of annihilations to some extent, if not negate it completely~\cite{Rapp:2000gy,Pan:2012ne,Seifert:2018bwl}.

\begin{figure}[t]
  \centering
  \includegraphics[width=.49\textwidth]{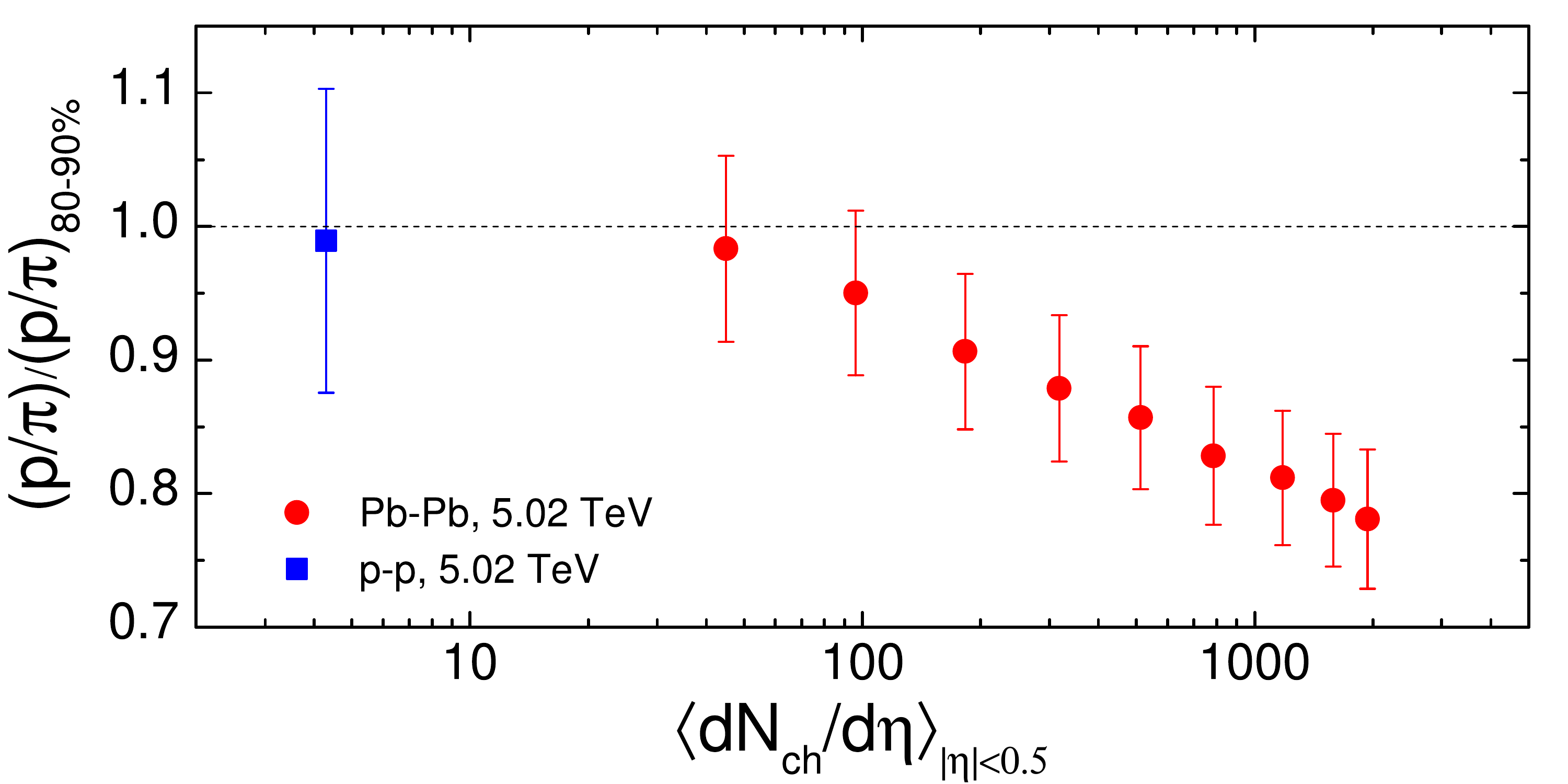}
  \caption{
  Centrality~(charged particle multiplicity) dependence of the $p/\pi$ ratio scaled by its value in peripheral (80-90\%) collisions in 5.02~TeV Pb-Pb collisions as measured by the ALICE Collaboration~\cite{ALICE:2019hno}. 
  }
  \label{fig:ppicentr}
\end{figure}

Thus, to which extent the proton yield may be modified in central collisions by baryon annihilation remains an open issue. 
Precision measurements of proton number fluctuations have recently been suggested to tackle this problem~\cite{Savchuk:2021aog}, as well as baryonic charge balance functions~\cite{Pratt:2022kvz}.
In the present work, we instead explore the centrality dependence of the $p/\pi$ ratio.
The uncertainties in the proton yield within SHM due to the modeling of hadronic interactions correspond to the evaluation of its chemical equilibrium abundance at a given temperature. Therefore, these uncertainties alone are not expected to generate any centrality dependence for the $p/\pi$ ratio, as long as the hadronization temperature is assumed to be centrality independent.
On the other hand, the hadronic phase is more relevant in central collisions compared to peripheral ones, as evidenced by centrality dependence of the kinetic freeze-out temperature~\cite{Abelev:2013vea,ALICE:2019hno} and resonance suppression~\cite{ALICE:2014jbq,Knospe:2015nva,ALICE:2018qdv,Motornenko:2019jha,ALICE:2019xyr}.
Baryon annihilation during a long-lived hadronic phase in central collisions can thus be expected to suppress the $p/\pi$ ratio relative to peripheral collisions. This effect is indeed observed in hadronic afterburner simulations at different centralities~\cite{Becattini:2014hla}.
Indications for this suppression are present in 2.76~TeV Pb-Pb data of the ALICE Collaboration~\cite{Abelev:2013vea}. However, it has been challenging to make definitive conclusions due to large systematic uncertainties in the data.
Recently, the ALICE Collaboration has published the $p/\pi$ data from the Pb-Pb run at 5.02~TeV~\cite{ALICE:2019hno}. 
These data have smaller error bars compared to 2.76~TeV. However, more importantly, the systematic uncertainties in the new data have been split into two contributions: (i) correlated and (ii) uncorrelated with centrality.
Using the much smaller uncorrelated uncertainty allows one to establish the suppression of the $p/\pi$ ratio in central collisions with a sizeable statistical significance.
Indeed, Fig.~\ref{fig:ppicentr} depicts the charged particle multiplicity dependence of the $p/\pi$ ratio at 5.02~TeV scaled by its value in peripheral~(80-90\%) collisions, where only the centrality-uncorrelated part of systematic uncertainties was used in the error propagation. 
Although the error bars still appear to be correlated with centrality, the results indicate the presence of statistically significant suppression of the $p/\pi$ ratio with multiplicity.
The largest suppression of the $p/\pi$ ratio is in 0-5\% collisions, with a suppression factor of $0.78 \pm 0.05$, with a significance of more than $4\sigma$.

In the present work we interpret the centrality dependence of the $p/\pi$ suppression as the effect of baryon annihilation in the hadronic phase. We also estimate the freeze-out temperature $T_{\rm ann}$ for the annihilation reactions from the experimental data.
To achieve this, we use the partial chemical equilibrium~(PCE) framework~\cite{Bebie:1991ij} to model the hadronic phase in heavy-ion collisions at the LHC~\cite{Motornenko:2019jha}. This framework is extended here to incorporate annihilation and regeneration reactions $N\bar{N} \leftrightarrow \mean{n_\pi} \, \pi$ involving (anti)nucleons and pions.

\paragraph{\bf Nucleon-antinucleon annihilation in partial chemical equilibrium.}

The PCE framework describes the gas of hadrons and resonances in partial chemical equilibrium, where all inelastic reactions are forbidden but elastic~(e.g. $\pi\pi \leftrightarrow \pi\pi$) as well as pseudo-elastic reactions involving short-lived resonances~(e.g. $\pi\pi \leftrightarrow \rho$, $\pi K \leftrightarrow K^*$, and $\pi N \leftrightarrow \Delta$) are equilibrated~\cite{Bebie:1991ij,Huovinen:2007xh}.
In this case, the total abundances of stable hadrons, including the feeddown from short-lived resonances, play the role of conserved quantities. 
For example, the total pion and nucleon numbers, which read $N^{\rm tot}_{\pi} = N_{\pi} + 2N_{\rho} + 3N_{\omega} + \ldots$ and $N^{\rm tot}_{N} = N_{N} + N_{\Delta} + N_{N^*}$, respectively, are conserved, thus one can introduce effective chemical potentials $\mu_\pi$ and $\mu_N$ that regulate their values. The same applies to kaons, stable hyperons, and long-lived resonances.
The chemical potentials of short-lived resonances are not independent but related to effective chemical potentials of their decay products through the condition of relative equilibrium of their decay and regeneration reactions, e.g., $\mu_\rho = 2 \mu_\pi$, $\mu_\Delta = \mu_N + \mu_\pi$, and so on.

Due to the isentropic nature of the fireball expansion in PCE~\cite{Bebie:1991ij}, all chemical potentials at a given temperature $T$ can be determined by solving the system of conservation equations
\eq{
\frac{n_{i}^{\rm tot}}{s} = \frac{n_i^{\rm ini}}{s^{\rm ini}}, \qquad i \in \pi, K, N, \ldots
}

Here $\frac{n_i^{\rm ini}}{s^{\rm ini}}$ are the initial value of the total stable hadron per entropy ratios, which in heavy-ion collisions correspond to the beginning of the hadronic phase.
The hadronic matter is assumed to be chemically equilibrated at the beginning of the hadronic phase. Thus, $\frac{n_i^{\rm ini}}{s^{\rm ini}}$ correspond to the values calculated in the statistical hadronization~(SHM) model.
The PCE framework has earlier been used to model the hadronic phase in hydrodynamic simulations~\cite{Huovinen:2007xh} and provides a reasonable description of resonance suppression~\cite{Motornenko:2019jha}. However, as inelastic reactions such as baryon annihilation are not allowed in the standard PCE framework, it requires modification.

Let us add reactions $N\bar{N} \leftrightarrow \npiNN \pi$, where $N \in p,n$ and $\pi \in \pi^+,\pi^-,\pi^0$, into the PCE framework. Typically $\npiNN = 5$~\cite{Dover:1992vj}, although the framework can incorporate also other values of $\npiNN$.
The total numbers $N_{N}$, $N_{\bar{N}}$, and $N_{\pi}$ are no longer conserved but instead a quantity,
\eq{
N^{\rm tot}_{\rm ann} = \frac{N_N^{\rm tot} + N^{\rm tot}_{\bar{N}}}{2} + \frac{N_\pi^{\rm tot}}{\npiNN},
}
is conserved as well as the net number of nucleons $N^{\rm net,tot}_N = N_N^{\rm tot} - N^{\rm tot}_{\bar{N}}$.
These two conservation equations are not sufficient to fix three chemical potentials, $\mu_N$, $\mu_{\bar{N}}$, and $\mu_\pi$.
Thus, an extra condition is required in order to close the system of equations.
The PCE framework is built on assuming relative chemical equilibrium of (pseudo-)elastic reactions, thus it is natural to assume in an extended PCE framework that the annihilation reactions $N\bar{N} \leftrightarrow \mean{n_\pi^{N\bar{N}}} \pi$ do also proceed in relative equilibrium. This implies the following relation for the chemical potentials,
\eq{\label{eq:chem}
\mu_N + \mu_{\bar{N}} = \mean{n_\pi^{N\bar{N}}} \, \mu_\pi,
}
which closes the system of equations.

Note that the relation~\eqref{eq:chem} assumes equilibrium of the reactions $N\bar{N} \leftrightarrow \mean{n_\pi^{N\bar{N}}} \pi$ during the hadronic phase.
Formally, this corresponds to an instantaneous annihilations equilibration time $\tau_{\rm eq}^{N\bar{N}} \to 0$. 
The validity of this assumption is questionable.
Generally, it is required that the $N\bar{N} \leftrightarrow \mean{n_\pi^{N\bar{N}}} \pi$ reaction rate is larger than fireball expansion rate to maintain equilibrium.
Alternatively, the equilibration time $\tau_{\rm eq}^{N\bar{N}}$ should be smaller than the duration of the hadronic phase.
A simple estimate for the equilibration time is $\tau_{\rm eq}^{N\bar{N}} = (\mean{\sigma_{N\bar{N}} v_{\rm rel}} n_B)^{-1}$ where $\mean{\sigma_{N\bar{N}} v_{\rm rel}} \sim 30-70$~mb~\cite{Dover:1992vj,Rapp:2000gy,Satarov:2013wga,Pan:2012ne} is the thermal-averaged cross section of the $p\bar{p}$ reaction, and $n_B \sim 0.03$~fm$^{-3}$ at $T = 160$~MeV~\cite{Vovchenko:2019pjl} is the number density of (anti)baryons, giving $\tau_{\rm eq}^{N\bar{N}} \sim 5-11$~fm/$c$.
This value is comparable to the hadronic phase lifetime of 4-8~fm/$c$ in central Pb-Pb collisions~\cite{Pan:2012ne,ALICE:2019xyr}, and thus it does indicate that the annihilations in the hadronic phase cannot be neglected.
However, it may also question the assumption that annihilation and regeneration reactions are close to equilibrium.
On the other hand, one can infer a smaller equilibration time of $\tau_{\rm eq}^{N\bar{N}} \sim 2-3$~fm/$c$ from Monte Carlo transport model simulations implementing $N\bar{N} \leftrightarrow 5\pi$ reactions through stochastic rates~\cite{Garcia-Montero:2021haa}.
Such a small $\tau_{\rm eq}^{N\bar{N}}$ would justify the equilibrium assumption, at least for the initial stages of the hadronic phase.
In the following, we thus adopt the generalized PCE framework and the associated equilibrium assumption but also discuss the possible corrections to this picture if this assumption is relaxed.

So far, we have only discussed the $N\bar{N}$ annihilations, which we have explicitly incorporated into the generalized PCE framework. Other baryons are also affected, however.
In particular, the condition~\eqref{eq:chem} implies the presence of baryon-antibaryon annihilation reactions involving all other non-strange baryons such as $\Delta$ and $N^*$.
Let us, for example, consider the lowest-lying $\Delta$ resonance.
Its chemical potential is $\mu_{\Delta(\bar{\Delta})} = \mu_{N(\bar{N})} + \mu_\pi$, reflecting the relative chemical equilibrium of $\Delta \leftrightarrow N \pi$ decays and regenerations.
Given Eq.~\eqref{eq:chem} one, therefore, has $\mu_\Delta + \mu_{\bar{N}} = (\npiNN + 1) \mu_\pi$ and $\mu_\Delta + \mu_{\bar{\Delta}} = (\npiNN + 2) \mu_\pi$ which implies relative chemical equilibrium of the annihilation reactions $\Delta \bar{N} \leftrightarrow (\npiNN + 1) \pi$ and $\Delta \bar{\Delta} \leftrightarrow (\npiNN + 2) \pi$, respectively.
The implication is that not only the primordial yield component of the total proton yield is affected by baryon annihilation, but also the feeddown contribution from resonance decays.

Partial chemical equilibrium has been implemented in the open source \texttt{Thermal-FIST} package~\cite{Vovchenko:2019pjl} since version 1.3, originally without $N\bar{N}$ annihilations.
In the present analysis, we use an extended version of the code that incorporates $N\bar{N} \leftrightarrow \mean{n_\pi^{N\bar{N}}} \pi$ reactions as described above.

\begin{figure}[t]
  \centering
  \includegraphics[width=.49\textwidth]{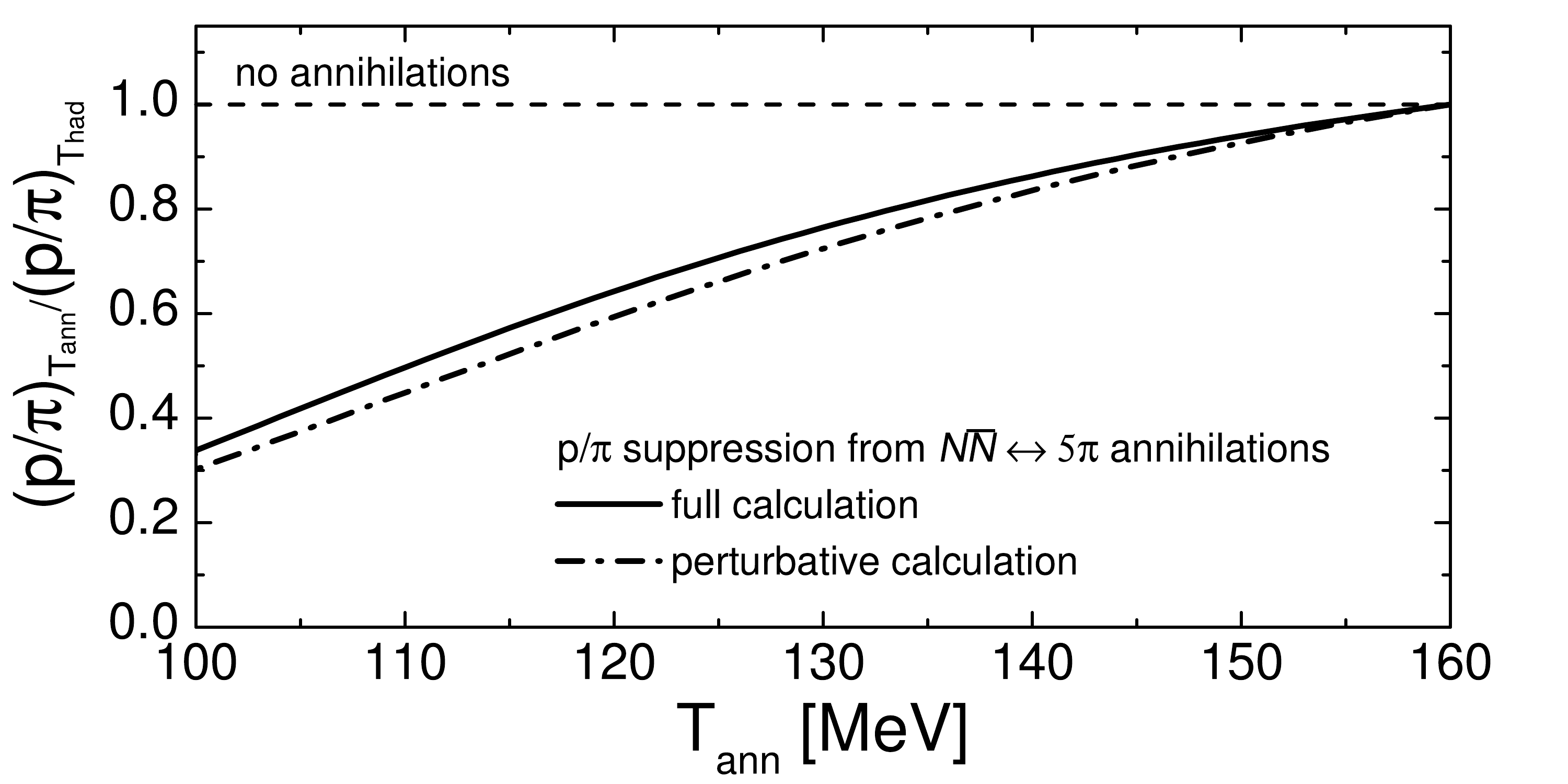}
  \caption{
  Temperature dependence of the $p/\pi$ ratio evaluated in the hadronic phase at $T = T_{\rm ann}$ in PCE model with $N\bar{N}$ annihilations relative to its value at the beginning of the hadronic phase~(hadronization) at $T = T_{\rm had} = 160$~MeV.
  The solid lines correspond to calculations within the generalized PCE framework that explicitly includes $N\bar{N} \leftrightarrow 5\pi$ annihilations, while the dashed-dotted line corresponds to a perturbative calculation on top of the standard PCE framework.
  }
  \label{fig:ppiPCE}
\end{figure}

At LHC energies, the treatment of annihilations can be simplified. 
First, due to the vanishing net baryon density one has $N_N^{\rm tot} = N^{\rm tot}_{\bar{N}}$ and thus $\mu_N = \mu_{\bar N} = \frac{1}{2} \npiNN \mu_\pi$.
Second, due to the fact that matter is meson-dominated~($p/\pi^+ \simeq 0.05$)~\cite{ALICE:2019nbs}, one can treat baryon annihilations perturbatively on top of the standard PCE description.
Namely, one neglects the change of pion number due to annihilations and evaluates the effective pion chemical potential $\mu_\pi$ in the standard PCE framework.
The effective chemical potentials of (anti)nucleons are then evaluated as $\mu_N = \mu_{\bar N} = \frac{1}{2} \npiNN \mu_\pi$ which are then used to calculate nucleon abundances.
While we use the complete generalized PCE framework in most of our numerical calculations, we also test the accuracy of the perturbative approach.

\paragraph{\bf Annihilations freeze-out from the ALICE data.}

Here we analyze the 5.02 TeV ALICE data on the centrality dependence of the $p/\pi$ ratio suppression~(Fig.~\ref{fig:ppicentr}) in the context of baryon annihilations in the hadronic phase.
We assume that, at each centrality, the hadronic phase starts with hadronization at $T_{\rm had} = 160$~MeV and expands in the state of partial chemical equilibrium which includes baryon annihilation reactions $N\bar{N} \leftrightarrow \npiNN \pi$ in relative chemical equilibrium.
We take $\npiNN = 5$ based on experimental data on $p\bar{p}$ reactions~\cite{Dover:1992vj}.
The $p/\pi$ ratio evaluated at each temperature includes feeddown contributions from all strong and electromagnetic decays.
The effect of annihilations is to decrease the $p/\pi$ ratio as the fireball cools and expands~(Fig.~\ref{fig:ppiPCE}).

\begin{figure}[t]
  \centering
  \includegraphics[width=.49\textwidth]{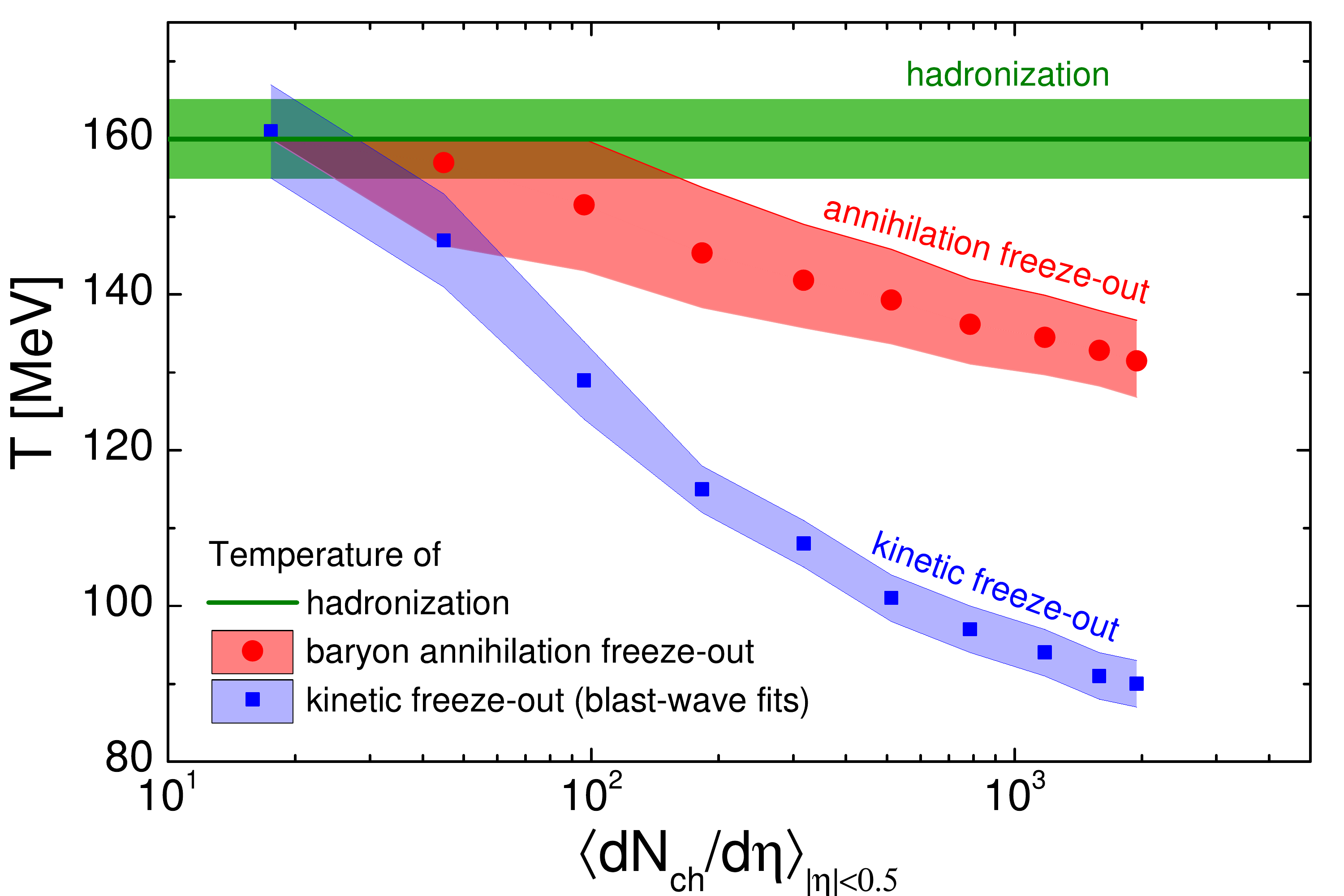}
  \caption{
  Annihilation freeze-out temperature~(red symbols with a band) extracted from the $p/\pi$ ratio in 5.02 TeV Pb-Pb collisions as a function of charged particle multiplicity~(centrality).
  Also shown are the hadronization temperature of $160\pm5$~MeV~(green line with a band) and the kinetic freeze-out temperature extracted from blast-wave fits~(blue symbols with a band)~\cite{ALICE:2019hno}.
  }
  \label{fig:Tann}
\end{figure}

Attributing the suppression of $p/\pi$ ratio in data to baryon annihilations, we estimate the annihilation freeze-out temperature $T_{\rm ann}$ at each centrality by matching the data~(Fig.~\ref{fig:ppicentr}) to the $p/\pi$ suppression predicted by the model~(Fig.~\ref{fig:ppiPCE}).
The corresponding results for $T_{\rm ann}$ are listed in Table~\ref{tab:Tann} and depicted in Fig.~\ref{fig:Tann} by the red band with symbols. The band width corresponds to the error propagation of the $p/\pi$ data in Fig.~\ref{fig:ppicentr}.
The resulting $T_{\rm ann}$ is a monotonically decreasing function of the charged particle multiplicity $\Nch$. 
In peripheral intervals, $\Nch \lesssim 100$, the annihilation freeze-out temperature is consistent with the hadronization temperature $T_{\rm ann} \simeq T_{\rm had} = 160$~MeV, indicating a short-lived hadronic phase and small relevance of the annihilation effects.
The relevance of baryon annihilations in more central collisions is evident, with the lowest $T_{\rm ann} = 132 \pm 5$~MeV value reached in 0-5\% central collisions~($\Nch = 1943 \pm 56$).
By construction, the effect of annihilation vanishes in the most peripheral bin, $80$-$90\%$. This assumption is supported by the fact that the $p/\pi$ ratio in peripheral Pb-Pb collision is consistent with the one measured in p-p collisions~(Fig.~\ref{fig:ppicentr}), where no annihilations are expected.
Nevertheless, one can relax this assumption in a more detailed study, which is left for future work. 

\begin{table}[!tbp]
     \centering
     \caption{Centrality dependence of the extracted baryon annihilation freeze-out temperature in Pb-Pb collisions at $\sNN = 5.02$~TeV.}
     \begin{tabular}{lccc} 
       \hline
       Centrality & $\mean{dN_{\rm ch}/d\eta}$    & $T_{\rm ann}$~[MeV]          \\
       \hline
       0$-$5\%    & 1943 $\pm$ 56  & 132 $\pm$ 5  \\
       5$-$10\%   & 1587 $\pm$ 47  & 133 $\pm$ 5  \\
       10$-$20\%  & 1180 $\pm$ 31  & 135 $\pm$ 5  \\
       20$-$30\%  & 786 $\pm$ 20   & 136 $\pm$ 6  \\
       30$-$40\%  & 512 $\pm$ 15   & 139 $\pm$ 6   \\
       40$-$50\%  & 318 $\pm$ 12   & 142 $\pm$ 7   \\
       50$-$60\%  & 183 $\pm$ 8    & 145 $\pm$ 8  \\
       60$-$70\%  & 96.3 $\pm$ 5.8 & 152 $\pm$ 8  \\
       70$-$80\%  & 44.9 $\pm$ 3.4 & $157^{+3}_{-11}$  \\
       80$-$90\%  & 17.5 $\pm$ 1.8 & 160  \\
       \hline
       \hline
     \end{tabular}
     \label{tab:Tann}
   \end{table}

It is instructive to compare the extracted $T_{\rm ann}$ values with the kinetic freeze-out temperatures $T_{\rm kin}$ that would correspond to the end of the hadronic phase.
The $T_{\rm kin}$ values are typically estimated from blast-wave fits to hadron $p_T$ spectra and their values for 5.02~TeV collisions~\cite{ALICE:2019hno} are shown in Fig.~\ref{fig:Tann} by the blue symbols with a band.
The $T_{\rm kin}$ values never exceed $T_{\rm ann}$ and are significantly below $T_{\rm ann}$ at all centralities apart from the two most peripheral bins.
This indicates an hierarchy $T_{\rm kin} < T_{\rm ann} < T_{\rm had}$ in (semi-)central collisions, implying that, in spite of their large cross sections, the annihilation reactions $N\bar{N} \leftrightarrow 5\pi$ freeze out earlier than the (pseudo-)elastic hadronic scatterings at the LHC energies.
This fact can be explained by the meson dominance of the hadronic matter created at the LHC, as indicated by low values of the measured baryon-to-meson ratios like $p/\pi \sim 0.05$.
This is also consistent with the expectation that baryon annihilation can only maintain equilibrium at early stages of the hadronic phase. 

As mentioned above, the generalized PCE framework assumes instantaneous equilibration time $\tau_{\rm eq}^{N\bar{N}} \to 0$ of $N\bar{N}$ annihilations. The actual equilibration time, however, is unlikely to be less than $\tau_{\rm eq}^{N\bar{N}} \sim 2-3$~fm/$c$ and is possibly even larger.
If one relaxes the $\tau_{\rm eq}^{N\bar{N}} \to 0$ assumption, it would follow that $p/\pi$ approaches the relative equilibrium value given by the generalized PCE at given $T_{\rm ann}$ only with some time delay, which is exacerbated by the fact that the system continues to expand and cool down.
This implies that the $p/\pi$ ratio predicted by the generalized PCE at $T_{\rm ann}$ shown in Fig.~\ref{fig:Tann} is likely reached at lower temperatures due to the fact that $N\bar{N} \leftrightarrow 5\pi$ reaction does not equilibrate instantaneously.
Therefore, the $T_{\rm ann}$ values shown Fig.~\ref{fig:Tann} should be regarded as the upper limit on the temperature values for the freeze-out of all nucleon number-changing reactions.

The result indicates that the measured proton~(and, to a much smaller extent, pion) yields should not be described by the chemical equilibrium SHM, as the yields are modified sizably by baryon annihilation in the hadronic phase.
Instead, the data on the $p/\pi$ ratio can be adjusted by modification factors in Fig.~\ref{fig:ppicentr}~(or computed through PCE framework if $T_{\rm ann}$ can be constrained in an independent way) to remove the effect of baryon annihilation and reconstruct the value at the hadronization stage.
The SHM fits can then be performed on these adjusted data to extract the hadronization temperature~(as opposed to chemical freeze-out temperature), as previously explored in Ref.~\cite{Becattini:2012xb} using modification factors from the UrQMD model.

\begin{figure}[t]
  \centering
  \includegraphics[width=.49\textwidth]{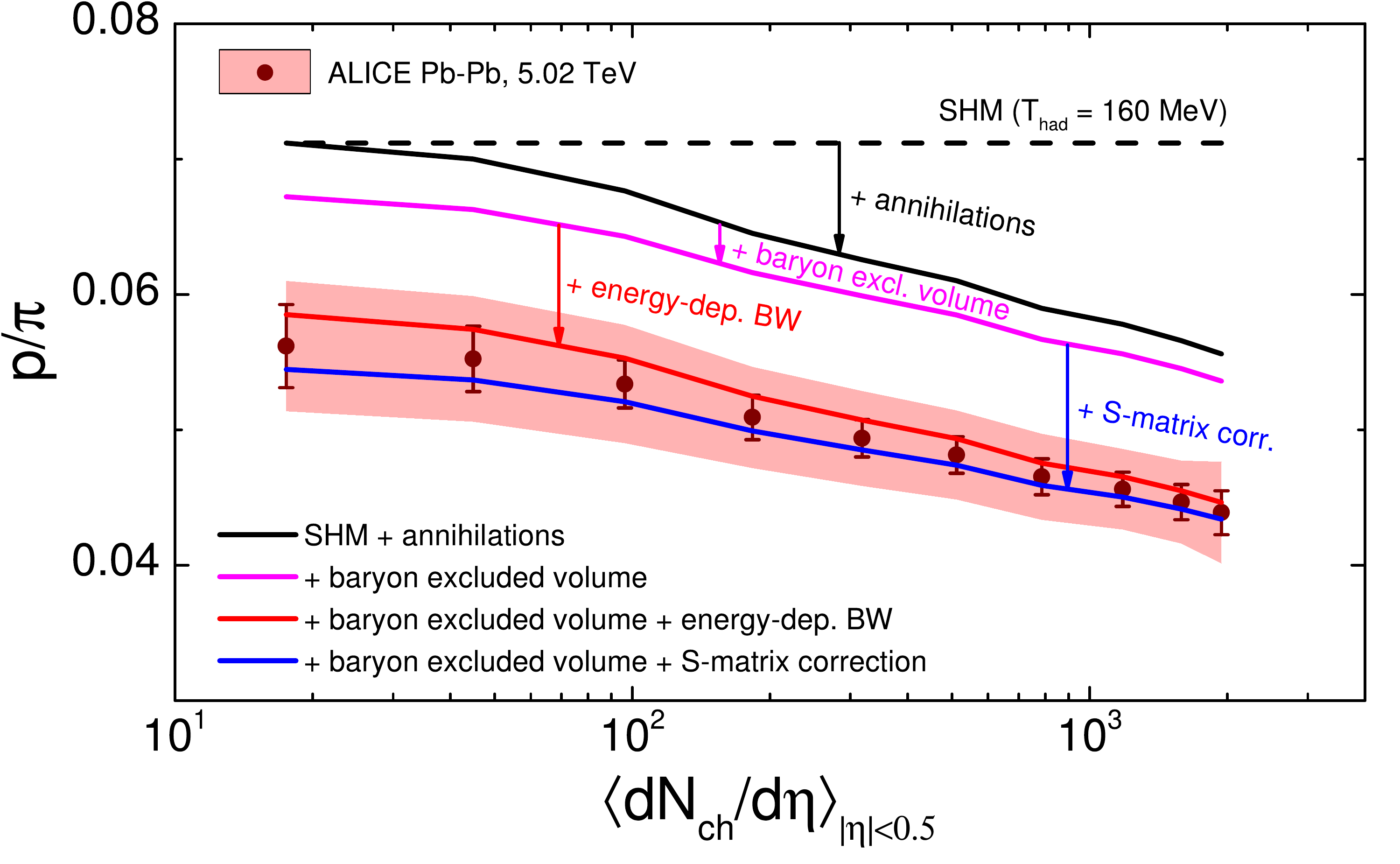}
  \caption{
  Multiplicity dependence of the $p/\pi$ ratio in Pb-Pb collisions at $\sNN = 5.02$~TeV calculated in the standard SHM model at $T_{\rm had} = 160$~MeV~(dashed black line), extended to include $N\bar{N} \leftrightarrow 5\pi$ annihilations~(solid black line), plus extended to include $\pi N$ interactions either via energy-dependent Breit-Wigner resonance widths~(solid red line)~\cite{Vovchenko:2018fmh} or S-matrix correction~(solid blue line)~\cite{Andronic:2018qqt}.
  The symbols show the experimental data of the ALICE Collaboration~\cite{ALICE:2019nbs}. The band and the error bars correspond to total and centrality-uncorrelated systematic uncertainties, respectively.
  }
  \label{fig:ppiall}
\end{figure}

\paragraph{\bf Other mechanisms affecting the $p/\pi$ ratio.}
Various mechanisms for modifying the proton yield in the SHM that are different from baryon annihilation have been explored in the literature as a possible explanation of the thermal proton yield anomaly in central collisions.
These include the excluded volume interactions~\cite{Alba:2016hwx}, finite resonance widths~\cite{Vovchenko:2018fmh}, or $S$-matrix corrections through $\pi N$ phase shifts~\cite{Andronic:2018qqt}.
These modifications do not predict any centrality dependence of the $p/\pi$ ratio and can thus be considered complementary to baryon annihilation.
Figure~\ref{fig:ppiall} shows how these various mechanisms influence the $p/\pi$ ratio.

When only annihilations are included~(solid black line), the model systematically overshoots the data at all centralities on a $3\sigma$ level.
This result indicates that baryon annihilation alone does not provide a complete resolution for the experiment's observed ``low'' $p/\pi$ ratio, but only its centrality trend.
It can thus be interesting to combine baryon annihilation with other mechanisms.
One such mechanism is short-range repulsion in the baryon-baryon interaction.
When modeled by means of an excluded volume prescription for baryons, with an excluded volume parameter $b = 1$~fm$^3$ fitted to lattice QCD data~\cite{Vovchenko:2017xad}, one obtains a 5\% reduction of the proton yield in SHM at $T = 160$~MeV~\cite{Vovchenko:2020kwg}.
Baryon excluded volume thus slightly improves the description of the $p/\pi$ ratio~(solid magenta line), but not sufficiently.
A more significant effect may come from reevaluating proton feeddown contributions from broad baryonic resonances such as $\Delta$ and $N^*$, which may be suppressed considerably relative to the standard SHM~\cite{Vovchenko:2018fmh,Andronic:2018qqt}.
When this effect is implemented through energy-dependent Breit-Wigner resonance widths~\cite{Vovchenko:2018fmh}, one obtains a much better description of the experimental data, with less than $1\sigma$ deviation at all centralities.
Similarly, when instead of Breit-Wigner one uses the $S$-matrix correction advocated in~\cite{Andronic:2018qqt} based on $\pi N$ scattering phase shifts, this leads to a similarly improved data description~(solid blue line). 
Note that the hadronization temperature is fixed to $T_{\rm had} = 160$~MeV throughout this analysis. 
We checked that using $T_{\rm had} = 155$~MeV yields generally similar results, although the data for the $p/\pi$ ratio tend to be slightly underestimated when all the discussed effects are included.
In a more detailed analysis, one can fit the value of $T_{\rm had}$ to experimental data.

Our analysis disregards the possibility of hyperon annihilation.
Although experimental constraints on these reactions are scarce, these reactions are expected to be relevant as well~\cite{Kapusta:2002pg}, and would thus suppress hyperon yields in central collisions~\cite{Becattini:2014hla,Stock:2018xaj}, qualitatively similar to the $p/\pi$ ratio. 
It would be interesting to return to this question once accurate data for the centrality dependence of hyperon-to-pion ratios become available.

\paragraph{\bf Effect on light nuclei production.}

\begin{figure*}[t]
  \centering
  \includegraphics[width=.32\textwidth]{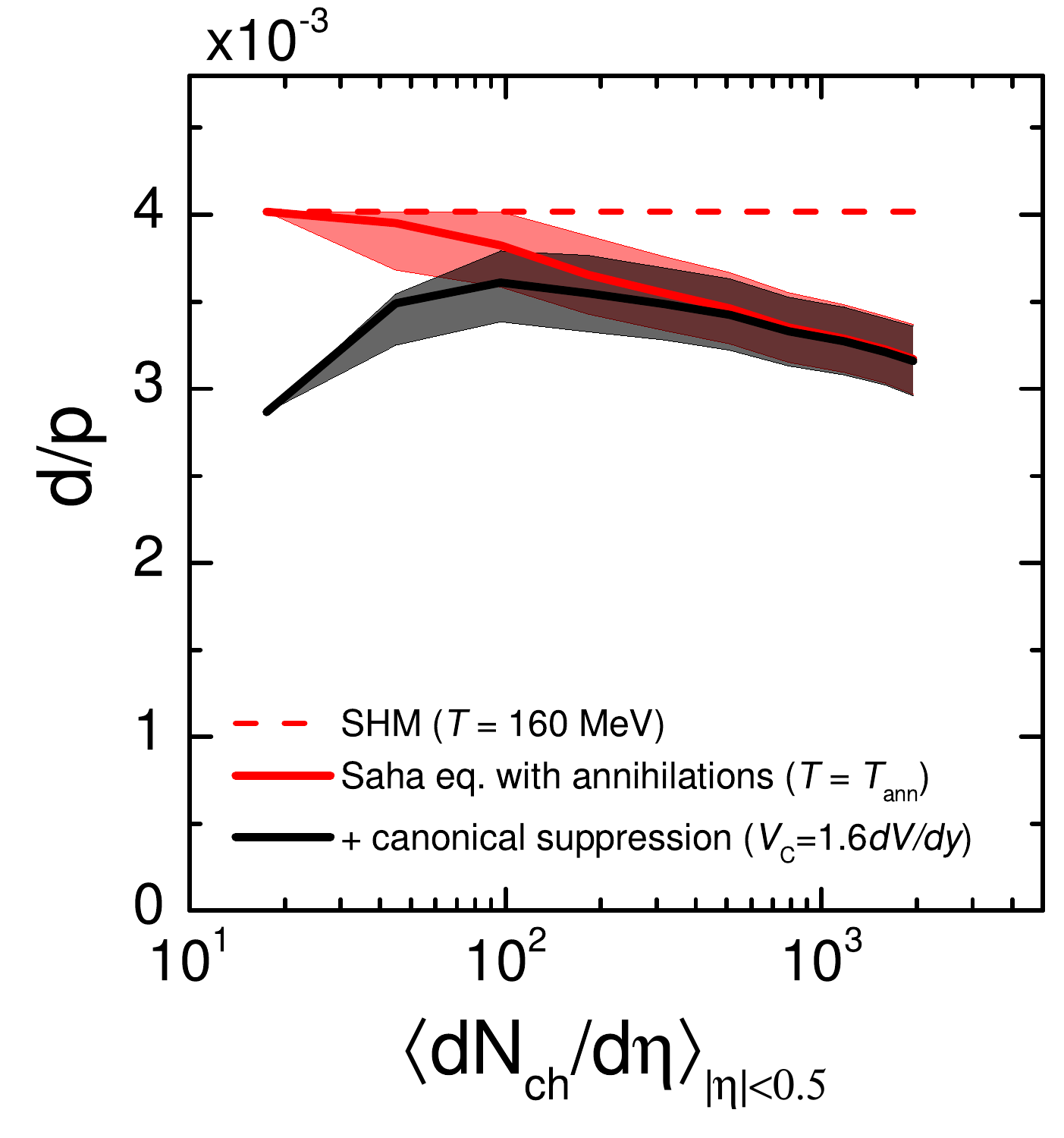}
  \includegraphics[width=.32\textwidth]{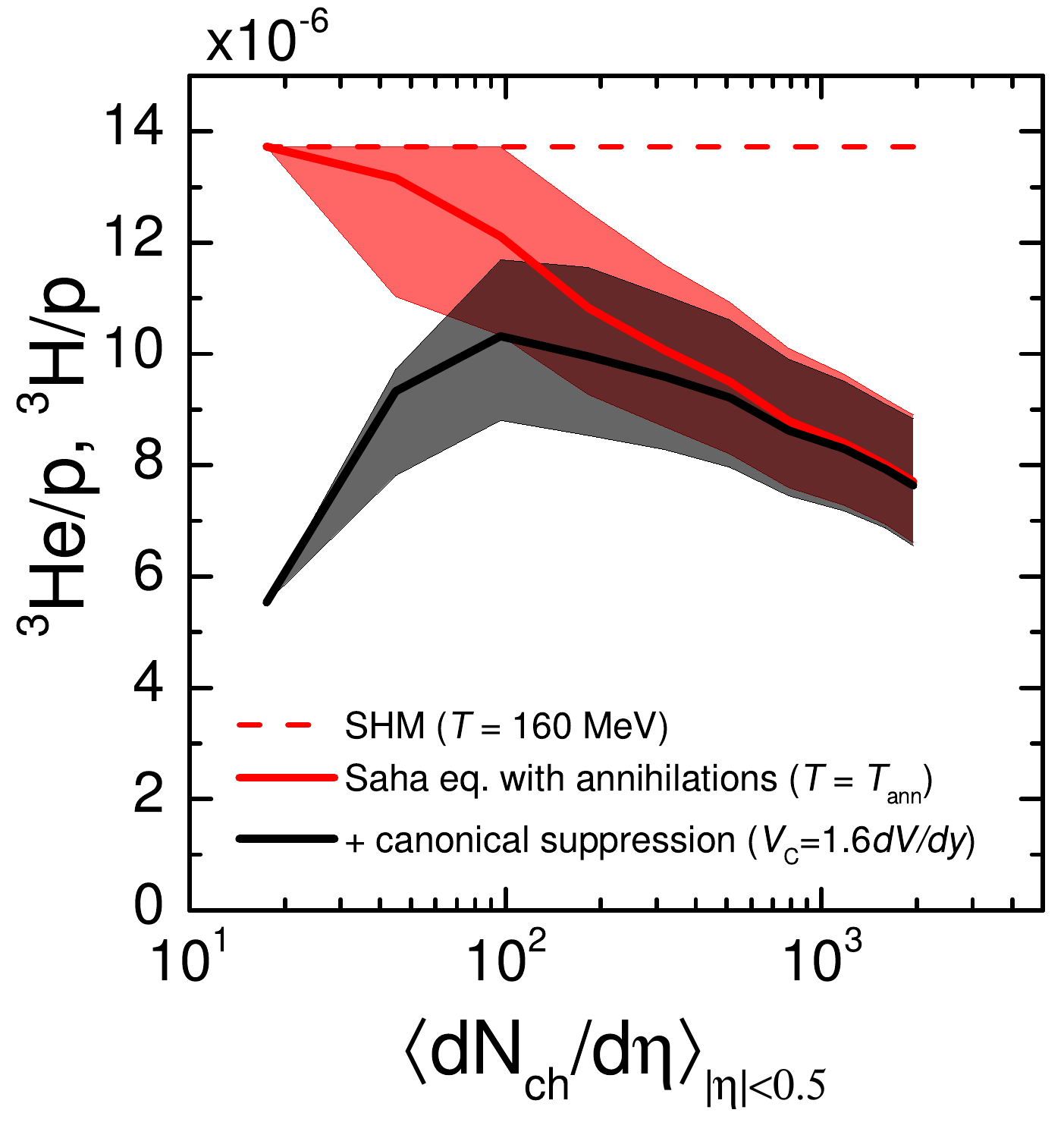}
  \includegraphics[width=.32\textwidth]{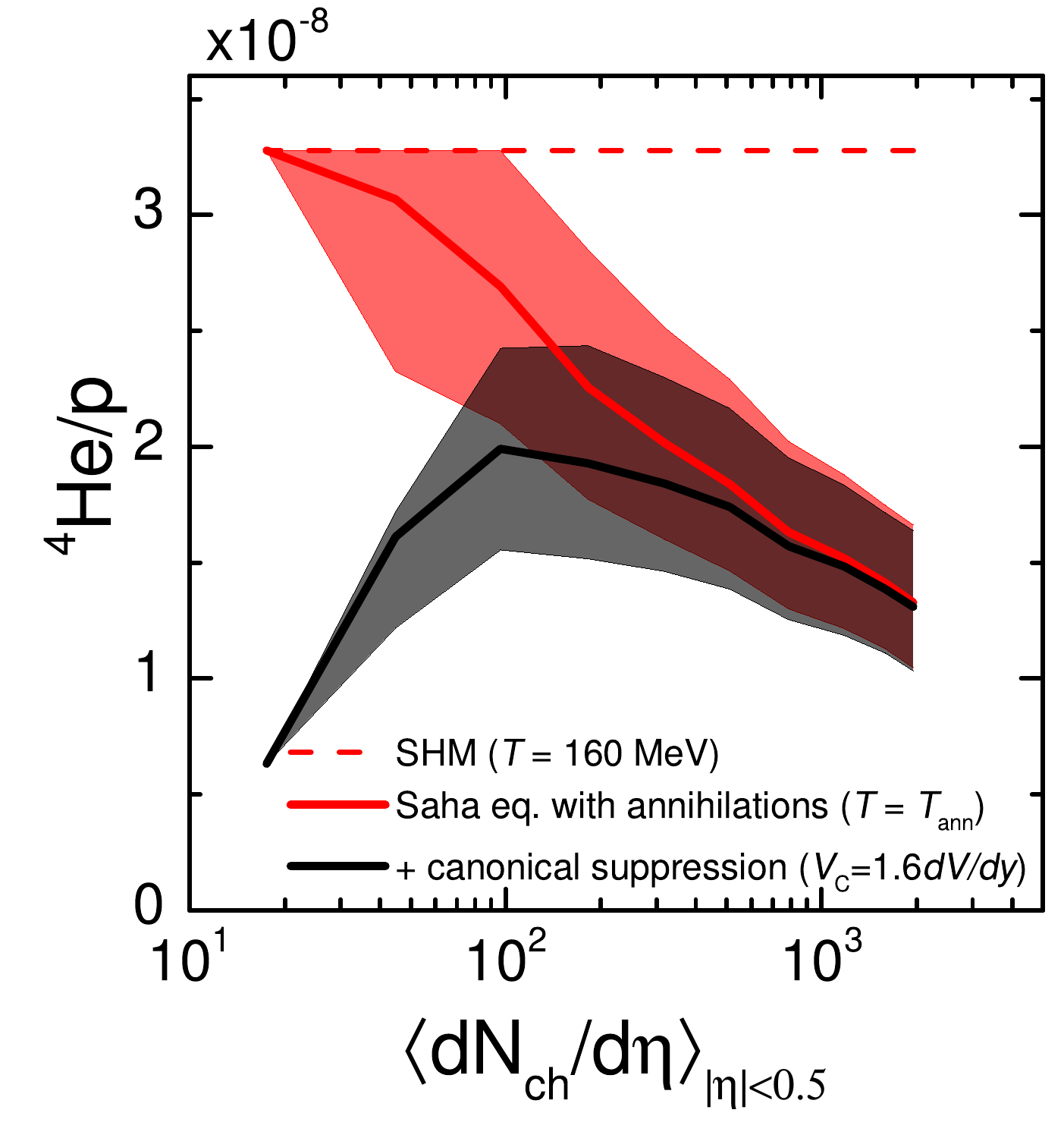}
  \caption{
  Multiplicity dependence of light nuclei to proton ratios calculated in the extended SHM~(Saha equation) with baryon annihilation~(red lines) and the additional effect of canonical suppression~(black lines). The calculations include the effect of finite resonance widths through energy-dependent Breit-Wigner prescription.
  }
  \label{fig:nuclei}
\end{figure*}

The results shown here are based on the assumption that the suppression of the $p/\pi$ ratio at various centralities relative to peripheral collisions can be entirely attributed to baryon annihilation.
It is thus instructive to consider experimental observables that could test this assumption.
Light nuclei are a natural candidate for such an observable, given that their constituents are nucleons.
In particular, if the annihilations suppress the nucleon yield by a factor $\gamma_N$, then, to leading order, the yield of a light nucleus $A$ would be suppressed by factor $\gamma_A \sim (\gamma_N)^A$.
For $\gamma_N \sim 0.8$ in central collisions this would imply, $\gamma_d \sim 0.64$ for deuterons, $\gamma_{^3\text{He}} \sim 0.51$ for $^3$He and $^3$H, and $\gamma_{^4\text{He}} \sim 0.41$ for $^4$He.
This type of suppression can be expected regardless of the exact mechanism for light nuclei production, as long as it assumes that nuclei are formed after the nucleon yields are frozen at $T = T_{\rm ann}$.

To make a quantitative estimate, we use the Saha equation approach~\cite{Vovchenko:2019aoz}, which allows one to compute light nuclei abundances in the hadronic phase.
Figure~\ref{fig:nuclei} depicts the ratios $\text{d}/p$, $^3$He/$p$, and $^4$He/$p$ as a function of charged particle multiplicity evaluated using the Saha equation at $T = T_{\rm ann}$~(red lines with bands). Here we include the effect of energy-dependent Breit-Wigner widths into our calculations to reproduce the right magnitude of proton yields.
The results show the expected suppression of light nuclei in central collisions, which becomes more prominent for heavier nuclei.
The effect of baryon annihilation obtained here is consistent with earlier studies employing UrQMD afterburner plus coalescence~\cite{Sombun:2018yqh,Reichert:2022mek}, or rate equations~\cite{Neidig:2021bal}.
Similar to the proton yield, the deuteron yield can be affected by hadronic interactions~\cite{Donigus:2022xrq}, the corresponding correction factors, however, are not expected to notably affect the centrality dependence.

One should note that light nuclei are also expected to be suppressed in small systems, as evidenced by the experimental data from pp and pA collisions at the LHC~\cite{ALICE:2019bnp,ALICE:2019fee,ALICE:2020foi}.
The canonical suppression from baryon conservation in the SHM approach~\cite{Vovchenko:2018fiy}, or the finite size of the emitting source relative to nuclear wave function in the coalescence approach~\cite{Sun:2018mqq}, have been discussed as possible mechanisms for this suppression. 
To illustrate this effect schematically, we apply canonical suppression factors from the canonical SHM~\cite{Vovchenko:2018fiy}, evaluated at $T = 160$~MeV and using canonical correlation volume $V_c = 1.6 \, dV/dy$~\cite{ALICE:2022amd}, to our calculations, this is shown by black lines with bands in Fig.~\ref{fig:nuclei}.
The canonical suppression in small systems leads to a non-monotonic multiplicity dependence of the light-nuclei-to-proton ratios, peaked at midcentral collisions.
These predictions can be tested with upcoming data from 5.02 TeV Pb-Pb run at the LHC where, similarly to the data on the $p/\pi$ ratio, the centrality-correlated part of the systematic uncertainty should be removed.

\paragraph{\bf Summary.}

We point out that the suppression of the $p/\pi$ ratio with charged particle multiplicity measured by the ALICE Collaboration in 5.02 TeV Pb-Pb collisions can be attributed to the presence of baryon annihilation in the hadronic phase.
By implementing the $B\bar{B} \leftrightarrow 5\pi$ reaction within a generalized partial chemical equilibrium framework, we estimate the annihilation freeze-out temperature at different centralities from the data, which is found to decrease with charged particle multiplicity from $T_{\rm ann} \simeq T_{\rm had} \simeq 160$~MeV in peripheral collisions to $T_{\rm ann} = 132 \pm 5$~MeV in 0-5\% most central collisions. This value is below the hadronization temperature but above the thermal (kinetic) freeze-out temperature of $T_{\rm kin} \sim 100$~MeV. The annihilation reactions thus remain relevant in the initial stage of the hadronic phase but freeze out before (pseudo-)elastic hadronic scatterings. This result indicates that proton yields should not be described by the standard chemical equilibrium SHM, unless the data are corrected for the proton yield modification in the hadronic phase.
One experimentally testable consequence of the annihilation picture is a suppression of the $\text{d}/p$, $^3$He$/p$, and $^4$He$/p$ ratios in central collisions of heavy ions, which calls for high precision measurements of these quantities as a function of charged-particle multiplicity.

\begin{acknowledgments}

\emph{Acknowledgments.} 
We thank Benjamin D\"onigus and Jan Steinheimer for reading the manuscript and useful comments.
We also thank Scott Pratt for fruitful discussions.
This work received support through the U.S. Department of Energy, 
Office of Science, Office of Nuclear Physics, under contract numbers 
DE-AC02-05CH11231231 and DE-FG02-00ER41132, and within the framework of the
Beam Energy Scan Theory (BEST) Topical Collaboration.
\end{acknowledgments}

\bibliography{NNann-PCE}

\end{document}